\documentclass[letterpaper,twocolumn,10pt]{article}
\usepackage{usenix2024_SOUPS}
\usepackage{amsmath}
\usepackage{array}
\usepackage[dvipsnames,table]{xcolor}
\usepackage{graphicx}
\usepackage{booktabs}
\usepackage{caption}
\usepackage{enumitem}
\usepackage{float}
\usepackage{hyperref}
\usepackage{multirow}
\usepackage{pgfplots}
\pgfplotsset{compat=1.18}
\usepackage{siunitx}
\usepackage{soul}
\usepackage{tabularx}
\usepackage{threeparttable}
\usepackage{comment}
\usepackage{xspace}
\usepackage{appendix}

\newcommand{\posts}{posts\xspace}

\newcommand{\pquote}[1]{\textit{``#1''}}
\newcommand{\header}[1]{\smallskip\noindent\textbf{#1.}}

\newcommand{\numInitialThreads}{1,310\xspace}
\newcommand{\numThreadsLLM}{80\xspace}
\newcommand{\numThreadsFP}{4\xspace}
\newcommand{\numThreads}{76\xspace}
\newcommand{\numPosts}{1,703\xspace}
\newcommand{\numThreadsP}{5.80\%\xspace}
\newcommand{\cutoff}{September 30, 2025\xspace}

\usepackage{fancyhdr}

\begin{document}
\date{}

\title{\Large \bf Like a Hammer, It Can Build, It Can Break: Large Language Model Uses, Perceptions, and Adoption in Cybersecurity Operations on Reddit
\thanks{This paper appears in the Proceedings of the Twenty-Second Symposium on Usable Privacy and Security (SOUPS) 2026.}
}

\def\plainauthor{Souradip Nath, Chih-Yi Huang, Aditi Ganapathi, Kashyap Thimmaraju, Jaron Mink, Gail-Joon Ahn}

\author{
{\rm Souradip Nath}$^{1}$,
{\rm Chih-Yi Huang}$^{1}$,
{\rm Aditi Ganapathi}$^{1}$,\\
{\rm Kashyap Thimmaraju}$^{2}$, 
{\rm Jaron Mink}$^{1}$,
{\rm Gail-Joon Ahn}$^{1}$
\\
$^{1}$Arizona State University
\quad $^{2}$Technische Universität Berlin
}

\maketitle

\begin{abstract}
Large language models (LLMs) have recently emerged as promising tools for augmenting Security Operations Center (SOC) workflows, with vendors increasingly marketing autonomous AI solutions for SOCs. 
However, there remains a limited empirical understanding of how such tools are used, perceived, and adopted by real-world security practitioners. 
To address this gap, we conduct a mixed-methods analysis of discussions in cybersecurity-focused forums to learn how a diverse group of practitioners use and perceive modern LLM tools for security operations. 
More specifically, we analyzed 892 posts between December 2022 and September 2025 from three cybersecurity-focused forums on Reddit, and, using a combination of qualitative coding and statistical analysis, examined how security practitioners discuss LLM tools across three dimensions: (1) their stated tools and use cases, (2) the perceived pros and cons of each tool across a set of critical factors, and (3) their adoption of such tools and the expected impacts on the cybersecurity industry and individual analysts.
Overall, our findings reveal nuanced patterns in LLM tools adoption, highlighting independent use of LLMs for low-risk, productivity-oriented tasks, alongside active interest around enterprise-grade, security-focused LLM platforms.
Although practitioners report meaningful gains in efficiency and effectiveness in LLM-assisted workflows, persistent issues with reliability, verification overheads, and security risks sharply constrain the autonomy granted to LLM tools. 
Based on these results, we also provide recommendations for developing and adopting LLM tools to ensure the security of organizations and the safety of cybersecurity practitioners.
\end{abstract}

\section{Introduction}
\label{sec:intro}

SOCs play a critical role in protecting organizations against an increasingly complex and fast-moving threat landscape.
They combine people, processes, and technology to support a wide range of defensive functions, including continuous monitoring, real-time detection, alert triage, and incident response~\cite{kokuluMatchedMismatchedSOCs2019,vielberth2020security,hofbauer2024blue}.
While many SOC processes have become increasingly automated through tools like SIEM, SOAR~\cite{hofbauer2024blue}, and traditional machine-learning (ML)–based techniques~\cite{sreelakshmi2024enhancing,andresini2021insomnia,capuano2022explainable,binbeshr2025rise,ghadermazi2024machine}, evidence from both industry and academia indicates that SOC teams remain under significant strain.
Industry reports show that SOCs receive thousands of alerts daily, with persistent false positives contributing to severe alert fatigue, stress, and burnout among analysts~\cite{tines-report-2023,vectra-report-2023}. 
Academic studies also highlight the limits of existing automation and the continued need for more effective decision support in the SOC~\cite{kokuluMatchedMismatchedSOCs2019,alahmadi202299,hassan2019nodoze}.

Against this backdrop, LLMs' generative reasoning capabilities have recently emerged as promising tools for augmenting SOC workflows~\cite{kramer2025integrating,singh2025llms,abuadbba2025promise}. 
Over the past years, research efforts have increasingly explored the application of LLM-powered automation to support a range of SOC tasks, from routine analysis to more advanced investigative workflows~\cite{lin2025ircopilot,tang2024nl2kql,muzammil2025towards,bisht2025improving,ismail2024threat}.
Capitalizing on this momentum, cybersecurity vendors have begun introducing autonomous solutions, such as Microsoft Copilot for Security~\cite{microsoft2024copilot} and CrowdStrike's Charlotte AI~\cite{crowdstrikecharlotteai}, to augment alert triage and incident response processes.

While these advances consistently highlight the transformative potential of LLM tools within SOCs~\cite{kshetri2025transforming}, understanding how these tools are used in practice and perceived by practitioners remains critical for assessing their real-world impact and for guiding the effective integration of LLMs into SOCs.
Prior to the rise of LLMs, researchers examined the perceptions and challenges of security practitioners regarding traditional ML–based security tools~\cite{mink2023everybody,oesch2020assessment}.
More recent work has shifted toward task-focused investigations of practitioners’ interactions with LLM tools, examining issues such as explainability, autonomy, trust, and human-AI collaboration within specific SOC contexts~\cite{roch2024navigating,rastogi2025survey,singh2025llms,kramer2025integrating}.
While these studies provide valuable insights into particular tools, tasks, or LLM capabilities, a broader understanding is still needed regarding how LLM tools are used across diverse SOC use cases, what motivating factors and barriers shape their adoption, and the expected impacts on the cybersecurity industry and individual analysts.

To address this gap, we turn to large-scale discourse in online cybersecurity forums.
Reddit has been widely used as a rich source of community discussion in security and privacy research~\cite{bouma2024honestly,oak2025victims,vakeva2025don,bouma2025scam}, hosting several large and active cybersecurity-focused communities, with the largest \texttt{r/cybersecurity} having over one million members as of January 2026.
In light of this, to capture natural, practitioner-driven discussions at scale around the uses, perceptions, and adoption of LLM tools in security operations, we conducted a {mixed-method} analysis of 892 \posts drawn from three active cybersecurity-focused forums on Reddit between December 2022 and September 2025 (inclusive), with the majority originating from \texttt{r/cybersecurity}. 
Based on this analysis, we seek to answer the following research questions:

\begin{enumerate}[start=1,leftmargin=*,noitemsep,label=\textbf{RQ\arabic*}]

    \item \textit{What LLM tools and use cases are mentioned in security practitioners’ discussions of SOC work?}\label{rq:AI-use}
    
    \item \textit{What benefits and drawbacks of LLM tools do security practitioners discuss in the context of SOC workflows?}\label{rq:AI-qualities}

    \item \textit{How do security practitioners discuss LLM adoption in SOCs and its potential implications for future practice?}\label{rq:AI-futures}
    
\end{enumerate}

Our findings reveal key patterns in how security practitioners discuss uses, perceptions, and adoption of LLM tools in cybersecurity operations:

\textit{First,} practitioner discussions reveal {a fragmented awareness of the rapidly expanding ecosystem of security-focused LLM platforms} with a small set of popular general-purpose models. Notably, tools such as ChatGPT and Microsoft Copilot, not explicitly designed for security operations, are referenced far more frequently than security-specific commercial tools, such as Security Copilot, Dropzone, or Intezer.

\textit{Second,}
LLM use cases span a broad range of operational SOC activities, with incident response and investigation support emerging as the most prevalent, followed by productivity-oriented tasks such as scripting and reporting. 
Furthermore, security-specific LLMs are more strongly associated with more autonomous triage and incident response workflows, whereas general-purpose LLMs are more commonly discussed for analyst-driven productivity tasks that afford greater control and easier verification.

\textit{Third,} practitioner discussions reveal clear gradients of autonomy: LLMs are most commonly used as decision support tools, less frequently embedded in human-in-the-loop triage pipelines, and only rarely granted fully autonomous control over end-to-end mitigation.

\textit{Fourth,} practitioners' perceptions of LLM tools are shaped by a balance of benefits and concerns across multiple factors. 
While many report positive experiences with the \emph{capabilities} and \emph{efficiency} of LLM tools in augmenting analyst workflows and reducing workload in specific contexts, these benefits are tempered by strong concerns related to \textit{insufficient reliability}, \textit{security and privacy risks}, \textit{limited autonomy}, and \textit{unjustifiable cost}. 
Overall, practitioners view LLM tools as \emph{promising yet insufficiently trustworthy} to warrant broad delegation of high-stakes SOC tasks without sustained human oversight.

\textit{Fifth,} practitioner discussions reveal a nuanced view of the workforce implications of LLMs within SOCs. 
While many anticipate that LLMs will reduce repetitive entry-level roles, practitioners overwhelmingly reject the idea of fully autonomous SOCs, emphasizing that higher-skill responsibilities, human oversight, governance, and accountability will remain fundamentally human-driven.

\header{Contributions}
Our study makes the following contributions:
\begin{itemize}[leftmargin=*, noitemsep]   
    \item We present the \textbf{first large-scale mixed-methods discourse analysis of practitioner discussions surrounding LLM adoption in SOCs}. Through a multi-dimensional analysis spanning tools, use cases, adoption patterns, practitioner opinions, and experiences, our study provides a broad empirical view of how practitioners perceive and operationalize LLMs within real-world security operations.

    \item We surface several \textbf{previously unexplored aspects}, including the fragmented ecosystem of security-specific LLM tools, emerging use cases such as knowledge discovery and learning, varying levels of autonomy in LLM-assisted workflows, practitioner curiosity and skepticism, and workforce implications in increasingly AI-augmented SOCs.

    \item Based on these findings, we derive \textbf{broader sociotechnical implications for LLM adoption in SOCs} and identify \textbf{research opportunities} surrounding trustworthy AI-assisted workflows, human-AI co-learning, and the long-term sustainability of the SOC workforce.
\end{itemize}
\section{Related Work and Background}
\label{sec:background}

\header{Human Factors in Security Operations}
Recent work has increasingly examined the human and organizational dimensions of security operations, 
using complementary methodologies, including qualitative interviews~\cite{alahmadi202299,kokuluMatchedMismatchedSOCs2019,thimmaraju2025human}, surveys~\cite{alahmadi202299,nepal2024burnout}, and anthropological analyses of analyst activities and tool usage~\cite{sundaramurthy2015human,tale-of-3,oesch2020assessment,yang2024true}.
Collectively, these studies highlight SOCs as complex sociotechnical environments and identify persistent operational issues, including alert fatigue, burnout, stress, and cognitive overload due to high alert volumes and false positives, along with mismatched priorities between analysts and organizational leadership.

Our work revisits these concerns in the context of LLM adoption within SOCs, examining both where LLMs may help alleviate existing burdens and what new challenges emerge as such systems become more widely adopted.

\header{Practitioner Perceptions of ML-based Security Tools}
Building on this foundation, subsequent research examined how ML-based security tools interact with operational SOC dynamics~\cite{mink2023everybody,oesch2020assessment}.
For instance, Mink et al.~\cite{mink2023everybody} conducted a qualitative study examining how practitioners perceive ML-driven detection systems and explanation mechanisms, identifying where such tools are considered effective compared to traditional rule-based approaches. 
Similarly, Oesch et al.~\cite{oesch2020assessment} analyzed analysts’ interactions with ML-assisted SOC tooling and highlighted challenges related to usability, analyst mental models, and the interpretability of ML-generated outputs.
Together, these studies demonstrate that the adoption of AI-assisted security tooling deeply depends on practitioners’ ability to understand, calibrate, and operationalize AI outputs.

Our work builds on these prior studies, particularly Mink et al.~\cite{mink2023everybody}, whose findings inform our initial codebook development~(\S~\ref{subsec:analysis}). 
However, our study extends the \emph{scope} and \emph{technological focus} by studying practitioners' perceptions of LLM-based tools, which introduce new forms of interaction, workflows, and capabilities that fundamentally reshape how AI systems are integrated into security operations.

\header{LLMs and Human-AI Collaboration in SOCs}
More recently, research has begun examining the role of LLMs and human-AI collaboration within SOCs~\cite{baruwal2024towards,mohsin2025unified,malatji2025augmented}, touching upon specific areas, such as autonomy, trust, interaction patterns, and explainability~\cite{roch2024navigating,rastogi2025survey,singh2025llms}.
For example, Roch et al.~\cite{roch2024navigating} conducted in-depth interviews with cybersecurity leaders to explore autonomy and the evolving division of labor between analysts and AI. 
Similarly, Rastogi et al.~\cite{rastogi2025survey} examined the role of explainability, highlighting how LLM-generated explanations influence trust and actionability in high-stakes SOC environments.
Other recent studies have explored task-specific uses of LLMs within SOCs, such as evaluations of LLM-assisted incident response summarization using real-world incidents~\cite{kramer2025integrating}.
Among these, Singh et al.~\cite{singh2025llms} is particularly relevant to our work, as it empirically examines analyst interactions with LLM systems in SOC settings. 
However, they focus on general-purpose LLMs such as ChatGPT, 
whereas our study additionally examines security-specific LLM platforms and more autonomous operational workflows.

Our work complements and extends this growing literature through its \emph{methodology}, \emph{scale}, and \emph{analytical breadth}. 
Rather than focusing on a specific tool, task, or AI capability, we conduct a large-scale multi-dimensional discourse analysis of practitioner discussions across online security communities. 
This approach captures focused yet candid conversations surrounding LLM adoption, operational use, organizational concerns, autonomy, trust, and perceived risks across a diverse range of tools and SOC use cases.
By analyzing practitioner discourse at scale, our work provides a broader view of how security practitioners collectively negotiate the emerging role of LLM systems within real-world security operations.

\header{Terminology}  
Next, we define the key terms related to data analytics and insights that are used throughout the paper.

\textbf{\emph{LLM Tools.} }
In this paper, the term `LLM tools' is used as an umbrella term to refer to a group of standalone tools, compound systems, or commercial platforms in which LLMs play a central role, enabling generative to agentic capabilities, such as reasoning, planning, and learning. 

\textbf{\emph{Practitioners.}}
In this paper, we refer to Reddit posters collectively as `security practitioners,' as the analyzed communities are cybersecurity-focused.
When available, we report more specific, self-identified roles (e.g., SOC manager) stated in posts or user flairs. 
Because these roles cannot be independently verified (\S~\ref{subsec:limitations}), we report them only when explicitly self-identified; otherwise, we use the term `practitioner.'

\textbf{\emph{Reddit Terms.} }
Throughout this paper, we use three Reddit-specific terms: \emph{subreddit}, \emph{thread}, and \emph{post}. 
Reddit is organized into topic-specific communities called `subreddits,'
within which users participate in discussion threads.
A `thread' refers to a user-initiated discussion consisting of a title, a conversation prompt, and all subsequent replies.
Following prior works~\cite{vakeva2025don,oak2025victims}, we use the term `post' to refer to any individual contribution within a thread, including both the original submission and any subsequent replies. 
\section{Methodology}
\label{sec:methodology}
To investigate how LLMs are used and perceived in practice, we perform a large-scale qualitative analysis across online cybersecurity forums discussing LLM tools.
We detail the ethical considerations for our IRB-approved study in Section~\ref{sec:ethics}.

\subsection{Data Collection}
\label{subsec:collection}
To accurately capture real-world security practitioners' uses and perceptions of LLM tools, we gathered and analyzed 1,703 \posts made across online cybersecurity forums.

\header{Discovering Subreddits}
Inspired by prior work analyzing Reddit data~\cite{bouma2024honestly,oak2025victims,vakeva2025don,bouma2025scam}, we selected a set of popular and active subreddits that enable discussions among security practitioners.
To discover security-centric discussions on emerging LLM tools, we began with informal keyword-based searches on Reddit to identify subreddits that yielded relevant practitioner discussions, and then expanded this set by including closely related, active communities with similar topical focus.
We selected active forums covering general cybersecurity discussions (\texttt{r/cybersecurity}, \texttt{r/ComputerSecurity}), cybersecurity leadership (\texttt{r/ciso}), blue-team and SOC operations (\texttt{r/blueteamsec}, \texttt{r/SIEM}), and cyberspace Q\&A (\texttt{r/CyberSecurityAdvice}). 
Furthermore, we read and adhered to each forum's posted community guidelines and avoided any communities that prohibited the use of forum data for academic research.
In total, we gathered a set of nine subreddits that spanned between $4.5$K--$1.4$M  users~(Table~\ref{tab:collection}). 

\begin{table}[thbp]
\centering
\small
\renewcommand{\arraystretch}{1.06}
\caption{\textbf{Collected Reddit Dataset} --  \textmd{{\small 
number of threads retrieved per subreddit, number and share of relevant threads, and the number (and share) of posts/comments coded as relevant.}}}
\newcolumntype{C}{>{\centering\arraybackslash}p{1cm}}
\resizebox{\columnwidth}{!}{
\begin{tabular}{@{}
                                >{\raggedright\arraybackslash}p{2.8cm} 
                                >{\raggedleft\arraybackslash}p{1.2cm}
                                >{\raggedleft\arraybackslash}p{1.6cm}
                                >{\raggedleft\arraybackslash}p{1.2cm}
                                >{\raggedleft\arraybackslash}p{1.8cm}@{}}
\toprule
\textbf{Subreddits} &
\textbf{Threads (Pulled)} &
\textbf{Threads (Relevant)} &
\textbf{Posts (Pulled)} &
\textbf{Posts (Relevant)} \\
\midrule
\texttt{r/cybersecurity}          & 830 & 69 (8.31\%)  & 1,665 & 863 (51.83\%) \\
\texttt{r/Information\_Security}  & 96  & 5 (5.21\%)   & 21    & 15  (71.43\%) \\
\texttt{r/ciso}                   & 25  & 2 (8.00\%)   & 17    & 14  (82.35\%) \\ 
\midrule
\texttt{r/blueteamsec}            & 187 & 0   & --    & --           \\
\texttt{r/CyberSecurityAdvice }   & 76  & 0   & --    & --           \\
\texttt{r/cybersecurity\_help}    & 36  & 0   & --    & --           \\
\texttt{r/ComputerSecurity }      & 35  & 0   & --    & --           \\
\texttt{r/SecurityBlueTeam}       & 15  & 0   & --    & --           \\
\texttt{r/SIEM}                   & 10  & 0   & --    & --           \\
\midrule
\textbf{Total} &
\textbf{\numInitialThreads} &
\textbf{\numThreads\;(\numThreadsP)} &
\textbf{\numPosts} &
\textbf{892 (52.38\%)} \\
\bottomrule
\end{tabular}
}

\label{tab:collection}
\end{table}

\header{Gathering Threads and Posts}
Within each subreddit, we looked at how LLMs were perceived and used by its members.
First, we performed a keyword-based search to collect and filter threads; with a set of $13$ AI-focused terms including `\texttt{SOC AI}', `\texttt{AI Security Operations}', and `\texttt{LLM in cybersecurity}'~(full list in~\cite{reddit2026github}), we retrieved \numInitialThreads threads on \cutoff, collecting the submission’s title, content, anonymized username, posting date, and number of posts. 
We further refined the set of collected threads using AI-assisted classification.  
Using \texttt{GPT-4.1-mini} API, we created a few-shot prompt to guide the LLM in performing a relevance classification task, supplying it with the thread title and the original post content~\cite{reddit2026github}. 
To establish reliability, the primary author manually reviewed a random sample of $100$ threads, labeled each as `relevant' or `irrelevant'. 
Out of $100$ threads, only $2$ were incorrectly coded, and no relevant threads were missed, yielding a high inter-rater reliability ($\kappa${=}0.96). 
Having found reliability, we then used the LLM to classify the remaining threads at scale, yielding \numThreadsLLM relevant and 1,230 irrelevant threads.
All relevant threads were also further manually reviewed, and \numThreadsFP were deemed irrelevant, giving us \numThreads relevant threads.  
We also randomly sampled $50$ threads classified as irrelevant, and manually verified that all the classifications were correct; these threads focused on topics outside the scope of this study, including frameworks for securing AI systems, cybersecurity career pathways, and interview preparation, or syntactically similar but unrelated topics, e.g., SoC (System-on-chip) Security. 

Within each thread, we retrieved all posts underneath it, producing a dataset of \numPosts posts as shown in Table~\ref{tab:collection}. 
These posts were then manually coded for relevance following our coding protocol~(\S~\ref{subsec:analysis}). Specifically, posts were double-coded in batches of 50 with iterative inter-rater reliability (IRR) checks and conflict resolution until a high level of agreement was achieved ($\alpha {=} 0.88$).
A post was deemed irrelevant if it had been deleted by the creator, removed by moderators or the platform, or if it did not pertain to our research questions.
Additionally, redundant replies from the same user that repeated already shared opinions without contributing new information were also labeled irrelevant.
Of the 1,703 posts coded, 892 were considered relevant (see Table~\ref{tab:collection}).

\header{Data Availability Statement}
\label{subsec:openscience}
Consistent with research practices involving Reddit data~\cite{bouma2024honestly,oak2025victims,vakeva2025don,bouma2025scam}, we will not publicly release the dataset to protect the privacy of Reddit posters. Instead, interested researchers may contact the author to request an archived copy.
To promote reproducibility, we provide the list of keywords and the few-shot prompt used in our data collection process in our replication package~\cite{reddit2026github}.

\subsection{Data Analysis}
\label{subsec:analysis}

To analyze online discussions, we employed a mix of qualitative and quantitative methods~\cite{mattei2025m} to gain insight into the discussions and statistically compare the prevalence of different LLM tools use cases and opinions.

\header{Qualitative Analysis of Posts}
To analyze the gathered \posts, we first conducted a hybrid coding methodology~\cite{azungah2018qualitative}.
To ensure proper interpretation, all posts were analyzed in the context of any posts that they were in reply to.
If any posts contained available external links, as in other works~\cite{bouma2024honestly}, we reviewed all linked content as well.

To construct the codebook, a single coder began with a set of codes informed by prior work~\cite{mink2023everybody}, from which initial use cases (alert triage, threat analysis, documentation) and AI factors (efficiency, cost, security) were derived.
This codebook was then expanded through inductive coding of 350 \posts to capture LLM-specific practitioner discussions, which introduced emergent use cases, such as scripting, query support, knowledge support, and training, as well as additional LLM factors, including capabilities, reliability, autonomy, and privacy. 
To assess IRR for this initial codebook, a second coder was introduced and familiarized with the codebook through co-coding of $25$ \posts.
After developing a shared understanding, the two coders then independently coded a new set of $50$ \posts, after which IRR was computed using Krippendorff’s alpha ($\alpha$)~\cite{hayes2007answering,mcdonald2019reliability} for every code. 
The authors then resolved coding disagreements and adjusted the codebook to clarify and refine definitions.
If at least one code did not obtain substantial agreement ($\alpha \geq 0.8$), this process was repeated.
After eight rounds of double coding, agreement for all codes was achieved, and the primary coder then independently coded the remaining 1,278 \posts.
The final codebook and resulting reliability are presented in 
our replication package~\cite{reddit2026github}.

Following codebook development, the primary coder, along with the full research team, conducted a reflexive thematic analysis~\cite{braun2021thematic}. 
Through data exploration and routine meetings with the research team, we distilled patterns and trends in the codes into large-scale themes that we present in our results.

\header{Quantitative Analysis of Codes}
Having established the reliability of codes, we also conducted quantitative analysis to measure and statistically compare their prevalence.

In particular, to assess the prevalence of certain LLM tools (\S~\ref{subsec:tools}) and use cases (\S~\ref{subsec:tasks}) discussed, we conducted pairwise comparisons using two-sample tests for equality of proportions~\cite{proportions}, which assume sufficiently large sample sizes for the asymptotic approximation underlying the chi-squared test statistic to be valid~\cite{prop-test}.
Also, to examine associations between categorical dimensions in our dataset, we conducted a series of chi-square tests of independence across multiple analyses, including relationships between tool categories and use cases (\S~\ref{sec:results-rq1}), LLM factors and opinions, tool categories and opinions (\S~\ref{sec:results-rq2}), and adoption stages over time~(\S~\ref{sec:results-rq3}). 
For statistically significant results, we conducted post hoc analyses using adjusted standardized residuals ($z$)~\cite{unit2020using} to identify disproportionately represented categories. 
To account for multiple comparisons, we applied the Bonferroni correction to the post hoc tests~\cite{weisstein2004bonferroni}.

\subsection{Limitations}
\label{subsec:limitations}
Similar to other works that analyze anonymous community forums~\cite{bouma2024honestly,oak2025victims,vakeva2025don,bouma2025scam}, our study shares a number of limitations that we account for when interpreting our results.
\textit{First}, our selection of forums and keywords, while carefully curated, may reflect biases and capture only a subset of relevant discussions about practitioners' experiences with LLM tools in SOCs. 
\textit{Second}, we rely on self-reported information provided by Reddit users. Although we report specific roles when explicitly stated in \posts, we cannot independently verify the roles, professional identities, or organizational contexts of the contributors, which may substantially influence how certain tools or factors are perceived and prioritized.
\textit{Third}, Reddit users represent only a subset, and possibly a biased section of the security community. 
Prior studies have noted that Reddit users tend to be more engaged with emerging technologies than the general practitioner population~\cite{klemmer2024using}.
As such, these discussions may reflect an upper bound on awareness of LLM tools rather than typical practice.

Despite these limitations, our work surfaces nuanced practitioner perspectives and timely insights into emerging LLM adoption trends in SOCs, which can be further validated through empirical and design-oriented methodologies.
\section{Tools and Use Cases in Practice (\ref{rq:AI-use})}
\label{sec:results-rq1}
We find that security practitioners discuss a wide range of LLM tools and uses within SOC workflows.
In particular, these discussions reflect a dominant mention of general-purpose LLMs alongside a fragmented ecosystem of commercial tools for security.
Moreover, LLM use cases span a diverse set of SOC activities, with discussions most heavily concentrated on incident response and triage, followed by scripting and reporting tasks, and less frequently on knowledge support, threat analysis, and training-related functions.

\begin{table}[thbp]
\centering
\caption{\textbf{Reported LLM Tools} -- \textmd{\small Categorized by purpose and origin. Note that individual posts may mention multiple tools.}}
\footnotesize
\renewcommand{\arraystretch}{1.05}
\resizebox{\columnwidth}{!}{%
\begin{tabular}{@{}p{6cm}%
                >{\raggedleft\arraybackslash}p{2.5cm}@{}}
\toprule 
{\textbf{Categories of LLM Tools}} &  \textbf{$\#$}  \textbf{(Out of 410)} \\ 
\midrule
\textbf{By Purpose} & \\
\quad General-Purpose LLM Tools                               & 248  (60.49\%) \\
\quad Security-Specific LLM Tools              & 180 (43.90\%) \\ 

\textbf{By Origin} & \\
\quad Commercially Available                        &  398 (97.07\%) \\
\quad In-House Built                             & 14 (3.42\%) \\ 

\bottomrule 
\end{tabular}
}

\label{tab:tools}
\end{table}
\begin{table*}[thbp]
\centering
\caption{\textbf{Reported LLM Tools Use Cases and Descriptions Across Tool Categories (General-Purpose, Security-Specific).}}
\footnotesize
\renewcommand{\arraystretch}{1.08}
\resizebox{\textwidth}{!}
{%

\begin{tabular}{@{}p{4.05cm}p{9.5cm}%
                >{\centering\arraybackslash}p{1.6cm}
                >{\centering\arraybackslash}p{1.6cm}@{}
                >{\centering\arraybackslash}p{1.0cm}@{}
                }

\toprule
{\textbf{Types of LLM Use Cases}} 
& \textbf{Description} 
& \textbf{\# General} & \textbf{\# Security}
& \textbf{$p$-value}
\\ 
\midrule
 Triage \& Incident Response ($n$=$139$)       
    & \textit{Support for alert triage, investigation, correlation, mitigation, and response workflows.}
    & \textbf{30 (40.54\%)}
    & \textbf{44 (59.46\%)}
    & \textbf{< 0.001*}
    \\
 Scripting \& Query Support ($n$=$88$)         
    & \textit{Generation and refinement of security scripts and detection queries.}
    & \textbf{51 (89.47\%)}
    & \textbf{6 (10.53\%)}
    & \textbf{< 0.01*}
    \\
 Reporting \& Documentation  ($n$=$84$)        
    & \textit{Drafting and summarizing investigation reports, policies, and threat briefings, etc.}
    & \textbf{43 (89.58\%)}
    & \textbf{5 (10.42\%)}
    & \textbf{< 0.05*}
    \\
 Threat Analysis ($n$=$54$)                  
    & \textit{Support for risk and vulnerability analysis, threat hunting, and modeling workflows.}
    & 16 (57.14\%)  
    & 12 (42.86\%)
    & 0.948
    \\ 
 Knowledge Support  ($n$=$53$)                
    & \textit{Concept explanation, information retrieval, document-based knowledge extraction.}
    & 31 (88.57\%)
    & 4 (11.43\%)
    & 0.119
    \\ 
Training, Compliance, Others  ($n$=$18$)   
    & \textit{Drafting training exercises, reviewing compliance materials, and other non-core tasks.}
    & 9 (81.82\%) 
    & 2 (18.18\%)
    & --
    \\ 

\bottomrule
\end{tabular}
}

\vspace{2pt}
\raggedright
\footnotesize{
Counts and percentages are reported row-wise and represent the relative distribution of tool types within each use case category. Statistical significance was assessed using a chi-square test of independence followed by Bonferroni-corrected post-hoc analysis of adjusted residuals. Rows denoting significant associations are bolded.}

\label{tab:usecase}
\end{table*}

\subsection{LLM Tools}
\label{subsec:tools}

Across 410 posts, practitioners discussed a wide range of LLM tools used within SOCs. 
To understand the tools used, we code each by whether it is 
(1) general-purpose, such as ChatGPT, or security-specific, such as Microsoft Security Copilot, and (2) commercially available or built in-house.

\header{General-Purpose LLMs Are Referenced More Than Security-Focused LLMs}
As shown in Table~\ref{tab:tools},  while not explicitly designed for security operations, 60.5\% ($n${=}$248$) of posts referenced general-purpose LLMs, such as ChatGPT, Microsoft Copilot, Claude, along with broader terms such as  ``LLMs,'' or ``Generative AI.'' 
Conversely, only 43.9\% ($n${=}$180$) of tools references mentioned security-specific tools, including Microsoft Security Copilot, Dropzone AI, and Intezer.
To evaluate whether these differences are significant, we ran a two-sample test for equality of proportions, and found that general LLM tools are significantly more likely to be discussed by security practitioners ($\chi^2(1)=21.94$, $p < .001$).
Across these discussions, practitioners described diverse integration patterns (\S~\ref{subsec:tasks}), alongside sharing hands-on reflections on their benefits and limitations (\S~\ref{sec:results-rq2}).

\header{A Long Tail of Security-Specific Tools Exists}
While general-purpose LLMs were centered around a few key players, a long list of security-specific tools was referenced (see Appendix~\ref{appndx:tools-list}).
Practitioners mentioned only \textit{nine} distinct general-purpose tools, with discussion dominated by ChatGPT ($n${=}$90$), followed by MS Copilot ($n${=}$30$), and others. 
In contrast, security-focused LLMs had a long list of 30 distinct commercial platforms, yet only four of these tools were mentioned more than five times: Security Copilot ($n${=}$40$), Dropzone AI ($n${=}$10$), Intezer ($n${=}$8$), and Cortex XSIAM ($n${=}$6$). 
The remaining appeared three times or fewer, with half of the 30 tools mentioned only once, indicating fragmented awareness across a rapidly expanding AI-for-cybersecurity vendor ecosystem.
Notably, half of these tools are marketed as ``AI SOC analysts'' (further discussed in \S~\ref{subsec:adoption}) positioned as autonomous assistants capable of performing Tier 1/2 tasks~\cite{kshetri2025transforming}.

\header{In-House Custom Workflows Are Rare, but Used}
In-house developments accounted for only 3.4\% ($n${=}$14$) of posts, compared to the commercially available solutions referenced in 97\% ($n${=}$398$) of posts.
Across this smaller subset, practitioners described building custom LLM workflows internally to augment security workflows.
For example, in a thread about LLMs for SIEM, one practitioner (P310) described leveraging open-source agent frameworks to enhance SIEM search workflows:
\pquote{I have developed custom Python scripts using open-source agent frameworks to feed SIEM query results into an agent to iteratively process the data and generate a consolidated review of the search output.}
This suggests that although off-the-shelf solutions are popular, such custom workflows may be valuable for closing integration gaps and supporting organization-specific security processes.

\subsection{Uses of LLM Tools in Security Operations}
\label{subsec:tasks}

As shown in Table~\ref{tab:usecase}, across 325 posts, LLM use cases fell into six broad categories: {triage and incident response}, {scripting support}, {reporting}, {threat analysis}, {knowledge support}, and {miscellaneous activities}. 
To understand the relative prevalence across use cases, we conducted pairwise two-sample tests of proportions (full statistical results in Appendix~\ref{appndx:stats}) and found that discussions surrounding triage and IR were significantly more prevalent than all other use cases, followed by scripting and documentation. 
Threat Analysis and knowledge support were discussed less frequently, but were more prominent than training, compliance, and other categories.

Furthermore, we conducted a cross-dimensional analysis linking the use cases with LLM tool categories (general-purpose or security-specific) and found that practitioners mentioned a specific tool across 60\% of use case discussions ($n${=}$193$). 
A chi-square test of independence (Appendix~\ref{appndx:stats}) revealed a significant association ($\chi^2(4)=58.289$, $p<0.01$); security-specific tools were more strongly associated with triage and incident response workflows, where applications tended to be more autonomous, whereas general-purpose tools were more commonly discussed for analyst-driven productivity tasks such as scripting, query support, and reporting. 
These findings suggest a functional distinction between productivity-oriented uses of general-purpose LLMs and operational automation expectations surrounding security-specific systems.

\header{Incident Response and Investigation Support}
Across all LLM cybersecurity use cases, 42.77\% ($n${=}$139$) of posts focused on improving or automating tasks relating to {incident response}, including triaging, investigating, responding, and mitigating alerts and security threats, and was thus mentioned significantly more often than all other reported use cases ($\chi^2(1) \geq 16.92$, all $p < .001$).
Practitioners described a varied set of ways for LLM tools to aid in reducing their workload and making investigations more effective.

\textit{LLM Investigation Buddies:}
Similar to common uses of LLMs in non-cybersecurity contexts, practitioners described using LLMs as decision support tools during investigations, operating in a {`think and assist'} mode that performs subtasks such as correlation, hypothesis generation, and sensemaking, while analysts retain full control over interpretation and final decision making.
As P452 shared: \pquote{During a complex investigation, I utilize LLMs as a companion, throwing ideas at them. I let them examine the data to help with correlation, and I handle the critical thinking.}
Similarly, P757 described providing contextual artifacts to LLMs for exploratory analysis, noting: \pquote{I provide logs, screenshots, or event timelines to LLMs to help me piece together what might be happening, either to validate my findings, or help me zoom in on a problem area.}

\textit{LLM-Driven Triaging:} 
Beyond providing investigative support, practitioners described using LLMs to partially automate alert triage through human-in-the-loop (HITL) workflows.
In these settings, HITL means workflows where LLMs may perform an initial analysis of incoming alerts, proposing likely scenarios and potential mitigation actions, but human operators review and approve or reject LLM decisions before any action is executed.
Such LLM-driven triage was most commonly discussed for high-volume `Tier-1' alerts, where rapid filtering decisions are required, though some accounts also described its use in more in-depth analytical contexts.
For instance, one L3 analyst (P725) described an LLM-driven pipeline that automatically extracts context from EDR alerts and presents a structured breakdown for analysis to ultimately review and take action on: \pquote{While we use it to augment the analysis and get more clarity on things, we do not allow it to take actions. Ultimately, trust but verify!}
Others, however, described allowing LLMs to handle and resolve low-impact alerts, escalating to humans only when necessary: \pquote{Our team has been using LLMs to assess high-noise/high-volume alerts that are low-payoff/low-impact. Only if there are any outliers, the alert is flagged for human evaluation} (P864).

\textit{Fully Autonomous LLM Mitigation:}
Lastly, a small fraction of posts ($n{=}5$) described {delegating entire classes of routine high-confidence tasks to fully autonomous LLM pipelines}.
Commonly integrated with SOAR systems, these workflows independently triage alerts, correlate incidents, enrich findings, and initiate actions with minimal human involvement (P013, P066, P247).
In a few cases, this autonomy extended into active responses; for instance, 
a CISO (P086) detailed how their LLM-powered correlation engine \pquote{determines incident severity and autonomously dispatches remediation tasks, like isolating endpoints, running full scans, or disabling suspicious accounts.}
These use cases demonstrate that, although limited, some organizations might be experimenting with {fully autonomous} workflows to scale high-volume alert workloads.

\header{Scripting \& Query Support}
Consistent with recent advances in LLM-based code generation~\cite{jiang2024survey,wang2023review,sobo2025evaluating}, 27.08\% ($n${=}$88$) of use cases described practitioners using LLMs to generate scripts and queries to augment their daily workflows.

\textit{Code Generation:}
Several \posts ($n${=}$46$) discussed using tools such as ChatGPT to draft scripts in Python, PowerShell, or Bash, typically to create boilerplate code or outlining logic that the analyst then adapts.
As P753 explained, \pquote{I find "write code to do XYZ in language A" prompts are useful for getting started with boilerplate code instead of writing it manually.}
Others mentioned leveraging LLMs to enhance or debug scripts they wrote themselves (P742).

{\textit{Query Support:}} 
Practitioners also mentioned LLM-assisted {query generation and refinement} ($n${=}$51$), particularly for SIEM and detection engineering workflows. 
These cases included assistance with crafting or troubleshooting KQL, SQL, or regex queries. 
As P309 noted, \pquote{I've explored a number of LLM chatbots to assist me fix SIEM queries, offer recommendations, or craft highly targeted queries for specific needs.}

\header{Reporting \& Documentation}
Across 25.85\% ($n${=}$84$) of mentioned tasks, practitioners described incorporating LLMs into their technical reporting and documentation workflows, supporting tasks ranging from \emph{general writing assistance} ($n${=}$20$) to SOC-specific activities such as \emph{policy and SOP development} ($n${=}$19$), \emph{investigation summaries} ($n${=}$13$), \emph{threat intelligence briefings} ($n${=}$11$),  and \emph{risk assessment reports} ($n${=}$9$).
As one mentioned using LLMs to offload time-intensive reporting tasks: \pquote{I use ChatGPT to ingest articles and generate summaries of CTI briefings. This allows me to save time and concentrate on threat hunting for any IOCs} (P823).

\header{Learning \& Knowledge Support}
Practitioners also described using LLMs as versatile knowledge-support tools ($n${=}$53$, 16.31\%), reflecting a broader shift in how analysts retrieve, learn, and navigate complex technical information.

\textit{Learning and Explanation:}
Across 20 posts, practitioners mentioned relying on LLMs to clarify obscure command-line behavior or provide context for vague error messages, highlighting the value of LLM-generated explanations for learning something new or complex. 
As P921 noted, \pquote{having an LLM explain something as if you were five may seem trivial, but can be incredibly helpful when approaching complex topics.} 

\textit{Information Retrieval:}
In 17 posts, LLMs were framed as a more efficient path from questions to actionable directions, often contrasted with a conventional Google search.
As P442, a threat hunter, described \pquote{the role of an analyst is to know the right question to get the answer. Google has been a part of our lives for many years. LLMs are simply far more efficient.} 

\textit{Knowledge Extraction:}
Finally, practitioners discussed utilizing LLMs to {{extract knowledge from documents}} ($n${=}$13$) to receive targeted responses.
For example, P740 described embedding internal materials into an LLM agent: \pquote{I created an agent loaded with our client policies, risk reports, and acronym lists to allow our analysts to ask questions or perform light analysis} (P740).

\header{Threat Analysis}
Around 16.62\% ($n${=}$54$) of posts discussed the use of LLMs to support analytical reasoning in activities such as risk and vulnerability assessment, threat modeling, and threat hunting.
For instance, P879 described using LLMs to help plan and review risk assessments, noting: \pquote{We have been augmenting LLMs extensively to our risk assessment planning to ensure completeness.}
Others reported using LLMs to sanity-check penetration testing findings, such as asking: \pquote{What description and CVSS score would you assign to my pentest discovery?} (P751), to save time by offloading the analytical thinking to the LLM.

\header{Training, Compliance, and Others}
A smaller subset of posts ($n${=}$18$, 5.54\%) described LLM use for miscellaneous tasks, including drafting cyber training exercises and developing scenarios to test defensive controls ($n${=}$6$), reviewing policies or guideline reports ($n${=}$6$), and other non-technical activities. 
As P823 noted: \pquote{I train junior analysts on IR, I use ChatGPT to generate training scenarios and practice questions.}
\section{Perceptions of LLM Tools (\ref{rq:AI-qualities})}
\label{sec:results-rq2}

\begin{table*}[thbp]
\centering

\caption{\textbf{Reported LLM Factors, Descriptions and Opinions.}}

\footnotesize

\renewcommand{\arraystretch}{1.08}

\begin{tabular}{p{2.1cm} p{9.2cm} c c c}
\toprule
\textbf{Factors} & \textbf{Description} & \textbf{\# Positive} & \textbf{\# Negative} 
& \textbf{$p$-value}
\\
\midrule

Capabilities
& \textit{Specific tasks and SOC use cases where LLM tools are perceived as helpful}
& \textbf{174 (64.21\%)}
& \textbf{97 (35.79\%)}
& \textbf{< 0.001*}
\\

Efficiency
& \textit{Time-related aspects, such as speed of analysis, deployment, and workload reduction}
& \textbf{54 (84.38\%)}
& \textbf{10 (15.62\%)}
& \textbf{< 0.001*}
\\

Reliability
& \textit{Trustworthiness, accuracy, and consistency of LLM-generated insights}
& \textbf{3 (5.17\%)}
& \textbf{55 (94.83\%)}
& \textbf{< 0.001*}
\\
Security \& Privacy
& \textit{Security of LLM tools, organizational data governance, and privacy considerations}
& \textbf{6 (11.54\%)}
& \textbf{46 (88.46\%)}
& \textbf{< 0.001*}
\\

Autonomy
& \textit{Ability of LLM tools to operate independently with minimal human supervision}
& \textbf{6 (12.77\%)}
& \textbf{41 (87.23\%)}
& \textbf{< 0.001*}
\\

Cost
& \textit{Financial aspects, such as operational costs and perceived return on investment}
& \textbf{5 (14.29\%)}
& \textbf{30 (85.71\%)}
& \textbf{< 0.001*}
\\

\bottomrule
\end{tabular}

\vspace{2pt}
\raggedright
\footnotesize{
Counts and percentages are reported row-wise and represent the relative proportion of opinions within each LLM factor.
Statistical significance was assessed using a chi-square test followed by Bonferroni-corrected post-hoc analysis of adjusted residuals. Rows denoting significant associations are bolded.}

\label{tab:factor-def}
\end{table*}

Across 406 \posts, practitioners shared opinions and experiences with LLM tools for cybersecurity operations.
Based on our analysis, we find that practitioners framed their experiences primarily around \textit{six} factors: {capabilities}, {efficiency}, {reliability}, {security and privacy}, {autonomy}, and {cost} (Table~\ref{tab:factor-def}).
To further examine whether sentiment toward LLM tools differed across the discussed factors, we conducted a chi-square test of independence followed by post-hoc analysis using adjusted residuals (full statistical results in Appendix~\ref{appndx:stats}). 
Our findings provide both statistical and qualitative evidence that, although practitioners are increasingly satisfied with the capabilities and efficiency of LLM tools, persistent concerns around reliability, autonomy, security, and cost temper their trust and limit their willingness to delegate sensitive tasks. 

\subsection{Capabilities}
\label{subsubsec:effectiveness-positive}

Practitioners’ discussions around LLMs' capabilities were significantly more positive than negative $(z=8.11, p < .001)$. 
Notably, nearly three-quarters of these posts ($n${=}$130$) also included specific use cases (\S~\ref{subsec:tasks}), highlighting that LLM-powered tools can effectively augment real-world SOC tasks.

\header{Better Contextualization and Visibility of Alerts}
Practitioners frequently described LLM tools as effective for {incident enrichment and surfacing low-visibility behaviors} ($n${=}$13$). 
While existing ML tools can correlate well-defined alerts, practitioners noted that LLM tools can effectively interpret a multitude of varied signals to better answer an investigation, which can also provide essential context for analyst interpretation.
For instance, reflecting on experiences with Agentic LLM tools such as Purple AI, one practitioner noted that, \pquote{LLMs are going to transform incident enrichment. Compared to scripting or googling, LLMs can rapidly synthesize key contextual details surrounding an incident} (P044).
Similarly, P038 noted that these tools dramatically reduce their workload, and
provide important context and remain easily accessible with natural language queries: 
\pquote{I use Agentic LLMs, and they're incredible. Not only does it turn half a million alerts/events a month into 1-3 relevant daily alerts, but also gives important context. During an investigation, I can ask, `Does this event fire every time a user logs in or is this a new alert?' or, `Did they get challenged for MFA?'... It's not just reducing my workload, it's finding things I couldn't see.}

\header{Improved Interpretability of Signals}
In addition to better contextualization of signals, practitioners ($n${=}$10$) also noted that LLMs made previously hard-to-understand signals readily interpretable.
In particular, practitioners emphasized how LLMs can translate verbose or opaque log data into human-readable narratives through querying, summarization, and explanations.
As P294 explained, \pquote{When you feed ChatGPT raw logs, it can assist in explaining what's happening in simple language. Windows events can be difficult to interpret due to the event IDs and combinations you need to memorize.} 

\header{Reduced False Positives}
Perhaps due to their effective use and analysis of many signals, practitioners also consistently framed LLM tools as effective in performing accurate classification decisions and automating alert triage and management pipelines ($n${=}$18$).
In particular, while concerns of high false negatives were notable in traditional alert classification systems~\cite{kokuluMatchedMismatchedSOCs2019, thimmaraju2025human,alahmadi202299}, and thought to be further exacerbated in traditional ML systems~\cite{mink2023everybody}, LLM tools were surprisingly perceived by analysts as broadly effective in detecting and re-emptively removing false positive alerts. 
For instance, P017 noted: \pquote{The `autonomous' aspect of LLM tools is most evident in their ability to eliminate high-confidence false positives that are not worth wasting an analyst's time.}

\header{Inabilities of LLMs}
Beyond the predominantly positive discussions of LLM capabilities, 97 posts (38.65\%) expressed reservations about their practical value in SOC contexts.
A recurring concern was whether the probabilistic nature of LLM tool outputs was fundamentally susceptible to incorrect conclusions, particularly around novel and unseen examples.
As P910 argued, \pquote{Cybersecurity mostly depends on dealing with outliers and anomalies... LLMs are awful at even comprehending these things, let alone acting upon.}
Skepticism also stemmed from firsthand experiences with performance breakdowns in complex or structured tasks, including failures to correctly interpret alert data or generate functional detection queries in complex languages, such as KQL.
Some posts emphasized systemic data challenges affecting LLM use at both training and inference stages.
As P263 emphasized that alert data already suffers from poor signal-to-noise ratios, suggesting that, \pquote{...with AI, the problem of garbage in, garbage out still persists.} 
Furthermore, these issues were compounded by vendors overpromising LLM tools' abilities, the perception that existing tools may be good enough (\S~\ref{subsec:future-caution}), and concerns that even outputs for tasks that LLMs can readily perform can also still suffer from hallucinations and inaccuracies (\S~\ref{subsubsec:reliability}).

\subsection{Efficiency}
\label{subsubsec:efficiency-positive}

Practitioners were significantly more likely to speak positively about the efficiency of LLM tools $(z=6.38, p < .001)$, 
frequently noting that LLM tools speed up their analysis. 

\header{LLM Tools Speed Up Analysis and Deployment}
Within core SOC workflows such as alert triage, investigation, and IR, practitioners frequently reported {measurable efficiency gains} from using LLM tools in their workflow.
For instance, an L3 analyst (P725) described how using an LLM tool in their workflows helped them automate their triaging and \pquote{reduced MTTT (mean-time-to-triage) from around 45 minutes to less than 2 minutes.} 
Similarly, the use of Agentic LLM tools drastically reduced how long investigations took by pre-emptively filtering and analyzing alerts they needed to respond to: \pquote{By delivering end-to-end investigations, these tools condense the investigation from "here’s 500 things to look at", to "here’s what happened and what you should probably do"} (P085).
Practitioners also noted that, compared with human analysts in particular, the {detection} and {response time} with LLM tools were often much faster for processing and resolving large volumes of alerts, with P870 noting: \pquote{AI can process massive volumes of data quickly and detect threats that would take a human analyst hours to identify.}
Furthermore, beyond the speed of the tool's operation, practitioners also noted that the time required to effectively deploy LLM tools was lower than that of others. This was discussed in both how fast analysts could learn to use (P002) and set up (P001) such tools. 

\header{LLM Tools Introduce Verification Overhead}
In contrast to their time-saving abilities, a small set of posts ($n${=}$10$) highlighted that LLM tools introduce new overhead as analysts may often need to correct, tune, or otherwise verify their outputs.
For example, P206 expressed frustration after their SOAR was replaced with an LLM tool-driven workflow: \pquote{We spend nearly four times the amount of effort correcting and changing LLM behavior as we do actually addressing incidents.} 
Similarly, P289 reflected after using LLMs for generating regular expressions for firewall rules: \pquote{I had to verify and correct the output, so did I end up saving time? Not sure.}
\subsection{Reliability}
\label{subsubsec:reliability}

When discussing the trustworthiness of LLMs' predictions, practitioners were significantly more likely to discuss negative aspects $(z=-6.77, p < .001)$ such as inconsistent behavior and false confidence in hallucinated answers.

\header{Hallucinations and Non-determinism Cause Concern} 
Most posts discussing reliability focused on LLM tools' tendencies to {hallucinate}~\cite{yao2023hallucination} ($n${=}$25$). 
Often, these were discovered through personal experiences; for instance, P171 noted that an LLM tool's confident but incorrect conclusion still makes them worry about relying on such tools: \pquote{Whenever I've tried LLMs for security work, it's produced pure garbage. It made up descriptions of an imaginary malware I named. I do not trust any of it.}
Beyond arriving at incorrect conclusions, practitioners noted that LLMs could also fabricate justifications and \pquote{hallucinate evidence to prove it} (P120). 
For many, this felt antithetical to the idea of cybersecurity; as noted by P924, \pquote{LLMs can be so confident in wrong answers... in the security space, these `small' mistakes may create cascading security risks by weakening multiple layers of defense.}
Furthermore, for others ($n${=}$4$), their concerns were not just based around incorrect answers, but the lack of determinism around LLM outputs.
Consistent with previous studies highlighting this issue~\cite{atil2024non,song2025good}, practitioners such as P058 noted: \pquote{The problem with implementing autonomous AI solutions for security is the unpredictable nature of the outputs.}
Practitioners also described how reliability can vary based on the specific task and its context; for instance, the particular language in which the code is generated: \pquote{When writing code, outcomes get increasingly unreliable as the code language gets more complex, like with PowerShell}~(P175).
\subsection{Privacy and Security}
\label{subsubsec:sec-privacy}

When discussing privacy and security risks of LLM tools, practitioners were significantly more likely to bring up negative viewpoints $(z=-5.41, p < .001)$. These concerns often focused on unintentional data leakage ($n${=}$31$), and the expanded attack surfaces LLMs may introduce ($n${=}$17$).

\header{Difficulties in Data Governance}
Practitioners frequently expressed concern that users may inadvertently enter organizational information into commercial LLMs, raising the risk that public models could retain or learn from sensitive inputs (e.g., P783, P827, P789).
As P294 cautioned: \pquote{LLMs may learn from the information you provide, be cautious about what you prompt; otherwise, sensitive organizational information may be inadvertently exposed.}
These concerns were particularly pronounced around integrated tools that require access to internal enterprise data.
As P863 shared, while their teams were excited by the success of early LLM tools, they remained \pquote{uncertain about the tool's access to internal data} and felt a tension between giving the tool access to be \pquote{useful without that access being a huge risk to themselves.}
Some practitioners questioned the necessity of such access and became worried that it may be misused or stolen: \pquote{It sounds like what all AI companies are trying to do: get as much data to train models that they can sell to other customers} (P885).

\header{LLMs Increase Attack Surfaces} 
Practitioners also raised concerns that LLM tools themselves become new attack surfaces.
In the excitement to produce and sell LLM tools for security, practitioners noted that security concerns of the tools themselves may not be prioritized: \pquote{Be careful, security with new technology often lags}~(P705).
Indeed, using simple techniques like jailbreaking~\cite{chu2025jailbreakradar} and prompt injections~\cite{liu2023prompt}, practitioners, like P926, became worried that LLM tools can be easily attacked: \pquote{I play with around LLMs a lot--it is alarmingly easy to get a model violate their guardrails.}
After breaking these boundaries, practitioners became worried that LLM tools themselves could be used to conduct attacks against their own company:  
\pquote{you could keep rephrasing open-ended questions to bypass its built-in safety restrictions} (P496).
\subsection{Autonomy}
\label{subsubsec:autonomy}

In discussing autonomous task execution without human supervision, practitioners were significantly more likely to hold negative perceptions $(z=-4.94, p < .001)$, noting that, despite vendor claims, LLM tools still required heavy oversight.

\header{While Improving, Human Oversight Is Still Needed}
Participants noted that while LLM tools are often marketed as {fully autonomous} and can act as {replacement analysts}, they were not often reliable enough to fully delegate tasks (P047, P257, P925). 
Most practitioners believed that current systems were better suited to decision support than to autonomous operations. 
As P047 mentioned, \pquote{I work with a wide range of tools that span the Gartner Quadrant; they are still deeply flawed and require human intervention and oversight.}
SOC work, in particular, was viewed as too nuanced, context-dependent, and constantly evolving for LLMs to fully handle (P011).
Thus, while LLMs were viewed as useful and in some cases essential, practitioners would not allow them to perform actions autonomously:
 \pquote{In my professional circle, we all agree that we will need to leverage AI, but only for advice, not for taking autonomous actions}~(P089).
 Instead, they recommended to others that they should \pquote{Use it like an intern. Allow it to gather information and raise its hand when it notices something, but don't let it touch anything critical} (P089).
\subsection{Cost}
\label{subsubsec:cost}

Practitioners were significantly more likely to be negative $(z=-4.02, p < .001)$ when discussing the cost of LLM tools.
While practitioners do not typically control organizational budgets, their negative perceptions around LLM costs reflect operational feasibility concerns within SOC workflows.

\header{Query Costs Add Up Fast}
Several practitioners noted that LLM inference costs~\cite{llm-cost} can become prohibitively expensive, raising concerns about the economics of scaling LLM-driven analysis to SOC-sized datasets. 
As P227 commented, \pquote{Each inference will cost a few cents, it might not be the most economical option for processing a large number of events.}
Moreover, some believed that privacy-concerned SOCs will need to invest heavily in computing infrastructure: \pquote{It's going to be so expensive to locally train and operate these high-quality LLMs--you will need data centers} (P895).

\header{Returns Are Limited}
Several participants were skeptical that, given this high cost, LLM tools were practical.
Costs were so high that practitioners, such as P790, questioned whether they could simply hire just as many workers with the same money: \pquote{Although Security Copilot has several interesting features, it is still not worth the price. Even minimal usage can be equivalent to one full-time employee's salary.}
Because of this, several participants, such as P362, believed that, \pquote{At this stage of its development, LLM tools may not provide enough value or a real return on investment.}
Other practitioners, such as P048 and P919, explicitly noted that they did not adopt LLM tools solely because of the cost.
\section{Adoption of LLM Tools (\ref{rq:AI-futures})}
\label{sec:results-rq3}

We now analyze how practitioners discuss adopting LLM tools within their operations and the concerns they hold.

\subsection{Trajectories of LLM Adoption}
\label{subsec:adoption}

By analyzing \posts ($n${=}$373$) for whether practitioners adopted LLM tools within their work, we found that 200 posts (53.62\%) reported actively using, 123 posts (32.98\%) were curious or in the process of evaluating LLM tools for adoption, and 50 posts (13.40\%) were not yet adopting or evaluating.

\begin{table}[t]
\centering
\caption{\textbf{Reported Adoption of LLM Tools Over Time.}}
\small
\renewcommand{\arraystretch}{1.15}
\resizebox{\columnwidth}{!}{%

\begin{tabular}{lccc|cc|cc}
\hline
                & \multicolumn{3}{c}{\textbf{Counts}} 
                & \multicolumn{2}{c}{\textbf{p1 vs p2}} 
                & \multicolumn{2}{c}{\textbf{p2 vs p3}} \\

\hline
\multicolumn{1}{l|}{\textbf{Adoption Stage}} 
                & \textbf{p1} 
                & \textbf{p2} 
                & \textbf{p3} 
                & \textbf{Delta} 
                & \textbf{$p$-val} 
                & \textbf{Delta} 
                & \textbf{$p$-val} \\
\hline
\multicolumn{1}{l|}{Evaluating }     & 20  & 16  & 87  & +4.40\%   & 1.000   & \textbf{-33.65\%} & \textbf{0.002*} \\

\multicolumn{1}{l|}{Using}           & 12  & 7   & 181 & -7.36\%   & 1.000   & \textbf{+31.09\%} & \textbf{0.013*} \\

\multicolumn{1}{l|}{Not Using}       & 3   & 3   & 44  & +2.97\%   & 1.000   & +2.56\%           & 1.000 \\
\hline
\end{tabular}%

}

\vspace{2pt}
\raggedright
\footnotesize{ 
\textbf{p1}: {Dec 2022--Oct 2023}, \textbf{p2}: {Nov 2023--Sep 2024}, \textbf{p3}: {Oct 2024--Aug 2025}
Pairwise comparisons were conducted using a chi-square test followed by Bonferroni-corrected post-hoc analysis of adjusted residuals. Bold values indicate statistically significant differences. Delta percentages are based on the proportions of adoption stages.}

\label{tab:adoption}
\end{table}

\header{LLM Adoption Shifts from Curiosity to Operational Use}
To examine whether practitioner attitudes toward LLM adoption changed over time, we divided the dataset timeline into three approximately equal intervals: (\textbf{p1}) \textbf{Dec 2022--Oct 2023}, (\textbf{p2}) \textbf{Nov 2023--Sep 2024}, and (\textbf{p3}) \textbf{Oct 2024--Aug 2025}, and compared the relative distribution of posts discussing each adoption stage (full statistical results in Appendix~\ref{appndx:stats}).

As shown in Table~\ref{tab:adoption}, between \textbf{p1} and \textbf{p2}, we observed no statistically significant changes across adoption stages ($p=1.00$).
Discussions in these earlier phases were similarly dominated by evaluating LLM tools, while active adoption remained comparatively limited, with only a few posts describing early operational use of ChatGPT ($n{=}6$), Microsoft Copilot ($n{=}2$), and Gemini in Google SecOps ($n{=}1$).
In contrast, statistically significant changes emerged between \textbf{p2} and \textbf{p3}, characterized by a significant increase in active operational use (from 26.92\% to 58.01\%, $p<0.05$) and a corresponding decline in exploratory discussions (from 61.54\% to 27.88\%, $p<0.01$), suggesting a shift from curiosity and experimentation toward concrete workflow integration.
Notably, discussions expressing non-adoption remained comparatively stable across periods ($p=1.00$), indicating that increasing operational use did not eliminate practitioner skepticism toward LLM tools. Instead, many cases of non-adoption stemmed from negative experiences during early evaluations (\S~\ref{subsec:future-caution}).

\header{General-Purpose LLMs Are Often Adopted, But Commercial Tools Drive Interest}
Across the posts describing active use of LLM tools, general-purpose LLMs were referenced most frequently and appeared in 103 posts (51.5\%), compared to only 37 posts (18.5\%) mentioning security-specific LLMs.
However, when looking at posts that discuss either actively evaluating or considering LLM use, security-focused LLM tools were referenced in 58 posts (47.16\%), more than double the frequency of general LLMs, which appeared in only 28 posts (22.76\%).
In general, this exploratory discussion came from several SOC managers (e.g., P001, P094, P305) actively exploring whether security-specific tools that promise near-automation of tasks were practical:
\pquote{We have started looking into AI SOC Analysts. Our team still spends a significant amount of time on L1/L2-type work, which should have been automated by now}~(P094).
Furthermore, practitioners sought community input on several related questions, including whether these LLM tools delivered value (P008, P055, P094, P860), {introduced overhead} (P055), {integrated well with existing technology} (P786), {compared favorably with traditional SOAR solutions} (P001, P008, P316, P411, P890), and {justified their cost} (P305, P890).

\header{Tools Are Often Adopted For Non-Core Security Operations} 
While broad discussions around use cases were often focused on core security operations (\S~\ref{subsec:tasks}), posts describing active adoption often focused on productivity-oriented tasks.
Of posts that described adopted uses for LLM tools,
28\% focused on scripting \& query support and 26\% on reporting and documentation; in comparison, 18\% discussed triage \& IR, and only 6\% discussed threat analysis.
This may indicate that while core operational tasks dominate overall discussions, active adoption is likely to concentrate on self-scoped tasks that afford greater analyst control and easier verification.

\subsection{Cautions against LLM Adoption}
\label{subsec:future-caution}

Despite reported adoption and curiosity, practitioner discussions highlighted recurring barriers that shaped how they evaluated the appropriateness of LLM tools for SOC.

\header{Inflated Vendor Promises}
Practitioners frequently interpreted LLM adoption through the lens of a long-standing history of overpromising technologies in cybersecurity, leading to skepticism toward marketing claims of “autonomous SOC” capabilities.
As P859 emphasized, \pquote{The marketing people will claim their product can do everything. However, without hands-on experiments, such products can be anything and everything at once, or nothing.}
Consistent with this view, across $20$ \posts, practitioners reported that they or their organizations had previously evaluated tools, such as Security Copilot, that did not result in adoption, primarily due to inefficient performance and unjustifiable costs.

\header{Sufficiency of Traditional Solutions}
Practitioner skepticism toward LLM adoption was also shaped by a perceived tendency to overuse, or \pquote{look for places to cram in AI} (P226).
As P463 argued that \pquote{many organizations lack the use case, scale, or resources to justify LLMs or agentic security solutions, yet they are forcing AI into workflows.}
Across 11 \posts, practitioners expressed a clear preference for traditional automation, emphasizing that many SOC tasks are already effectively addressed by existing approaches:
\pquote{It's like bringing a tank to a knife fight when using LLMs to deterministic security problems. Stick to traditional automation, it’s simpler, cheaper, and gets the job done} (P460).

\header{Organizational Restrictions and Policies}
Another cluster of narratives ($n${=}$10$) highlighted that organizations often restrict the use of public LLMs primarily due to {data-loss prevention} or {confidentiality concerns} (\S~\ref{subsubsec:sec-privacy}).
As P922 explained, \pquote{Since people continued to carelessly pour company data into public LLMs, we incorporated them into our company block list.} 
One practitioner also described broader uncertainty within organizations about how to assess the risks associated with LLMs, driven by the lack of established frameworks for evaluating these tools.
As P784 noted, \pquote{Since Gen AI is so new, nobody even understands how to discuss it…the CISO is unsure what the security risks are, the Chief Risk Officer doesn’t know how to characterize them on the risk matrix.}

\header{Anti-LLM Sentiment}
Lastly, a smaller subset of posts ($n${=}$9$) reflected a clear reluctance toward LLM adoption, rooted in personal or ideological opposition to the technology.
These \posts were rarely supported by detailed reasoning; instead, they expressed dismissiveness through statements such as \pquote{No, I have a brain} (P768), and \pquote{If people could stop talking about AI, I would pay for it} (P005). 
P840, however, articulated inhibition grounded in the broader implications of LLMs for the SOC workforce (further discussed in \S~\ref{subsec:future-jobs}). 
As they explained, \pquote{Organizations are eager to replace analysts with AI, making it difficult to be enthusiastic about technologies that may ultimately reduce the need for human roles.}

\section{Discussion}
\label{sec:discussion}

Next, we reflect on our findings and outline future research directions to understand the nuances of LLM adoption, design trustworthy systems, and sustain workforce development.

\header{LLM Adoption is Shaped by Control and Commitment}
Our findings suggest that LLM adoption in SOCs is not monolithic but rather a layered process that differs across tools and stakeholders.
General-purpose LLMs are often adopted by analysts through independent experimentation on self-scoped tasks, whereas much of the curiosity surrounding security-specific enterprise tools is expressed by self-identified decision-makers, reflecting organizational priorities~(\S~\ref{subsec:adoption}).
This split helps explain why SOCs may exhibit high reported adoption of general LLMs while still showing friction around enterprise adoption.
LLMs afford analysts greater control and reversibility, allowing them to selectively apply the tools to low-risk tasks without deep integration or organizational commitment. 
In contrast, adopting enterprise security-focused platforms is a procurement and governance-level decision~\cite{opdenbusch2025we, schinagl2020we} that often entails broader data access and tighter coupling with existing SOC tooling.

\emph{Research Opportunity.}
While recent work has begun to explore the adoption of LLMs for specific SOC tasks~\cite{kramer2025integrating,singh2025llms}, future research should examine adoption as a \emph{multi-stakeholder phenomenon}. 
Prior work has already identified inherent mismatches between how managers and analysts evaluate SOC work~\cite{kokuluMatchedMismatchedSOCs2019}; adopting a stakeholder-specific lens is therefore critical to understand the divergent priorities that shape adoption decisions. 
Future work should also investigate how informal \emph{analyst-level LLM use can be secured under governed deployments}, such as~\cite{syros2025saga}, without sacrificing the low-friction interaction patterns that make LLMs inherently valuable.

\header{Reliability is the Hard Ceiling for Autonomy}
Our findings indicate that practitioners’ reluctance to grant autonomy to LLM tools is not rooted in abstract distrust, but in hands-on experience with unreliable system behavior.
Reliability issues (\S~\ref{subsubsec:reliability}) necessitate manual verification before LLM-generated outputs can be applied, directly constraining the autonomy these systems can be granted.
When LLM judgments cannot be confidently trusted, delegation becomes a source of risk rather than benefit, echoing prior work linking low trust in automation to reversion to manual control~\cite{lee2004trust}. 

\emph{Research Opportunity.} 
This framing highlights the need for future research that treats \emph{trust and autonomy as interdependent properties}, and examines how LLM systems communicate uncertainty through \emph{reliability metrics} that allow analysts to assess the reliability of LLM tools in situ.
Prior work shows that generic uncertainty disclaimers, such as, ``this model may make mistakes,'' have minimal impact on the credibility threshold in high-stakes expert settings~\cite{kiyak2025chatgpt}.
Reliability metrics are therefore intended not to unconditionally increase trust, but to provide task-specific, actionable signals (e.g., via external hallucination detectors) helping practitioners decide when to act on LLM output and when to fall back to deterministic approaches, as a necessary step toward more trustworthy integration of LLMs within SOC workflows.

\header{The SOC Workforce Development Crisis}
Our findings surface an important tension with the implications for the long-term sustainability of the SOC workforce. 
Practitioners widely expect LLMs to reduce or replace entry-level responsibilities, while analysts are increasingly positioned as reviewers and governors of LLM-mediated workflows that presuppose substantial domain expertise (\S~\ref{subsec:future-jobs}).
However, prior research emphasized that such expertise develops incrementally through hands-on operational exposure~\cite{dawson2018future, goodall2009developing, baker2016striving}.
This creates a circular dependency: effective oversight of LLMs requires experienced analysts, yet the experiential learning pathways that produce such expertise through entry-level tasks are now being automated.
As a result, questions emerge about how future analysts will acquire the experiential knowledge needed to critically evaluate LLM decisions.

\emph{Research Opportunity.} 
Sustaining expertise in LLM-augmented SOCs may require \emph{rethinking training as a process of co-learning} between humans and LLMs. 
While machines have traditionally been learning from humans, recent work has begun to formalize co-learning paradigms through interaction, feedback, and shared problem-solving~\cite{lu2025we, huang2019human}. 
Early industry efforts are beginning to explore this space; COACH by Dropzone~\cite{coach} is an LLM-powered security mentor that provides junior analysts with real-time investigative guidance~\cite{dropzone-train}. 
Recent academic work also explored integrating LLM-powered tutoring into cybersecurity training environments~\cite{nelson2025sensai,wang2025cybermentor,chhetri2024exploring}.
Future research can explore how co-learning approaches can be grounded in cyber workforce development principles~\cite{baker2016striving}, while supporting LLM-mediated skill development.
\section{Conclusion}
\label{sec:conclusion}

In this paper, we present a large-scale mixed-methods analysis of practitioner discussions on Reddit, providing a multi-dimensional view of how LLM tools are used, perceived, and adopted within security operations.
Our findings reveal that while practitioners increasingly view LLMs as valuable for augmenting SOC workflows, particularly for self-scoped productivity tasks, concerns surrounding reliability, governance, and workforce sustainability continue to constrain broader operational delegation.
Together, these findings highlight that the future of LLM adoption in SOCs is fundamentally sociotechnical, shaped by organizational constraints, practitioner experiences, and long-term workforce considerations.
\section*{Ethics Statement}
\label{sec:ethics}

This study does not involve direct interactions with human subjects, as it relies exclusively on publicly available data~\cite{pritchard2001searching,sannon2022privacy}.
Nevertheless, we consulted the Institutional Review Board (IRB) at the leading institution and obtained an IRB exemption.
That said, we actively recognize the ethical implications of analyzing online public discourse and took steps to mitigate potential harms,
adhering to the ethical standards set in prior published works~\cite{bouma2024honestly,oak2025victims,vakeva2025don,bouma2025scam}.
\textit{First}, we acknowledge that Reddit users may not have anticipated their posts being used for research, and that neither community members nor moderators explicitly consented to such use. 
To address this, we followed best practices for Reddit data collection and limited our corpus to publicly accessible subreddits that do not prohibit research use under their terms of service.
Unlike some subreddits that explicitly restrict academic research due to the sensitivity of their content~\cite{rDrugsSubreddit-2026}, the communities we studied impose no such restrictions.
\textit{Second}, although the analyzed data is likely not sensitive, we implemented additional safeguards to protect user privacy. 
All identifying metadata (e.g., usernames and organization names) were removed prior to analysis, and contributors are referenced using anonymized identifiers (e.g., \texttt{PXXX}). 
To further reduce the risk of re-identification through reverse search, all excerpts included in this paper were carefully paraphrased in accordance with established ethical guidelines~\cite{fiesler2024remember,reagle2022disguising}.

\header{LLM Usage Considerations}
During the preparation of this manuscript, LLMs were used solely for editorial assistance, including language refinement and editing. 
All generated content was reviewed by the authors, who take full responsibility for the manuscript's accuracy, originality, and integrity.

\clearpage
\bibliographystyle{unsrt}
\bibliography{references}

\clearpage
\begin{appendices}

\section{Overview of Forum Threads and Posts}
\label{sec:results}
 
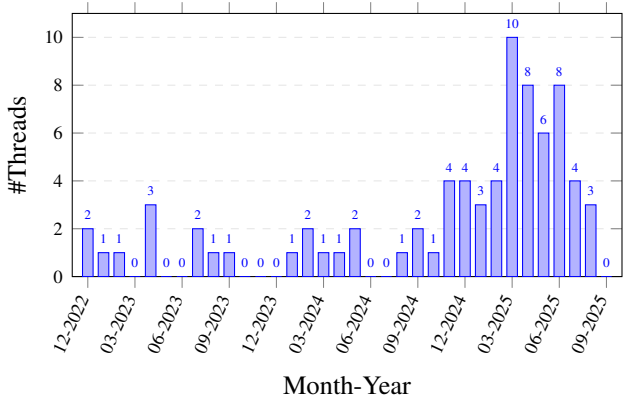
\begin{figure}[tbh]
\centering
\begin{tikzpicture}
\begin{axis}[
    ybar,
    width=0.50\textwidth,
    height=0.6\columnwidth,
    bar width=4pt,
    ymin=0,
    ylabel={\#Threads},
    ytick = {0,2,4,6,8,10},
    symbolic x coords={
    12-2022,01-2023,02-2023,03-2023,04-2023,05-2023,06-2023,07-2023,08-2023,09-2023,
    10-2023,11-2023,12-2023,01-2024,02-2024,03-2024,04-2024,05-2024,06-2024,07-2024,08-2024,
    09-2024,10-2024,11-2024,12-2024,01-2025,02-2025,03-2025,04-2025,05-2025,06-2025,07-2025,08-2025,09-2025
    },
    xtick={
    12-2022,03-2023,06-2023,09-2023,12-2023,03-2024,06-2024,09-2024,12-2024,03-2025,06-2025,09-2025
    },
    xticklabel style={rotate=65, anchor=east, font=\scriptsize},
    yticklabel style={font=\scriptsize},
    ymajorgrids,
    grid style={dashed,opacity=0.4},
    enlarge x limits=0.03,
    nodes near coords,
    nodes near coords style={font=\tiny, rotate=0},
]
\addplot coordinates {
  (12-2022,2) (01-2023,1) (02-2023,1) (03-2023,0) (04-2023,3) (05-2023,0)
  (06-2023,0) (07-2023,2) (08-2023,1) (09-2023,1) (10-2023,0) (11-2023,0) (12-2023,0)
  (01-2024,1) (02-2024,2) (03-2024,1) (04-2024,1) (05-2024,2) (06-2024,0) (07-2024,0) (08-2024,1)
  (09-2024,2) (10-2024,1) (11-2024,4) (12-2024,4)
  (01-2025,3) (02-2025,4) (03-2025,10) (04-2025,8) (05-2025,6) (06-2025,8) (07-2025,4) (08-2025,3) (09-2025,0)
};
\end{axis}
\end{tikzpicture}
\caption{\textbf{Number of Threads in Our Dataset Over Time.}}
\label{fig:thread-freq}
\end{figure}

The threads in our dataset came from three subreddits:
\texttt{r/cybersecurity} ($n${=}$69$), \texttt{r/Information\_Security} ($n${=}$5$), and \texttt{r/ciso} ($n${=}$2$). 
These relevant threads were posted between December 2022 and August 2025 (Figure~\ref{fig:thread-freq}).
We can see that conversations around LLM tools have increased over time, with 75\% of \posts made within the last year of data collection (Sep 2024--Aug 2025).
This aligns with the recent emergence and broader visibility of SOC-focused LLM tools around mid-2024, including Microsoft Copilot for Security~\cite{microsoft2024copilot}, Prophet Security’s AI SOC Analyst~\cite{prophet2024}, and the introduction of Gemini in Google SecOps~\cite{gemini2024}.

Relevant threads contained an average of 22 ($\pm 32.8$) \posts, ranging from 1–193 \posts.
Across these 76 threads, we analyzed 892 (52.38\%) relevant posts.
As shown in Table~\ref{tab:post-distribution}, these relevant \posts touched on our topics of inquiry: 
Tools~($n${=}$410$), Use Cases~($n${=}$325$), Opinions~($n${=}$406$), LLM Factors~($n${=}$459$), LLM Adoption~($n${=}$373$), and Vision for the Future~($n${=}$276$).

\section{Additional Results and Discussion}
\label{appndx:results}

\subsection{Commercial LLM Tools}
\label{appndx:tools-list}
Table~\ref{tab:tools-list} provides a comprehensive list of all commercial LLM tools, along with their purpose (general-purpose or security-specific), observed frequencies, and brief descriptions, mentioned in the practitioner discussions.

\subsection{Statistical Results}
\label{appndx:stats}

\begin{table}[t]
\centering
\footnotesize
\caption{\textbf{Distribution of Discussion Topics} -- \textmd{\small Identified across 892 relevant posts. Individual posts often addressed multiple themes.}}
\resizebox{\columnwidth}{!}
{

\begin{tabular}{@{}p{4.5cm}%
                >{\raggedleft\arraybackslash}p{3.6cm}@{}}
\toprule
\textbf{Topics of Discussion} 
    & \textbf{\# Total Posts (Out of 892)} \\ 
\midrule

\textbf{LLM Uses} (\S~\ref{sec:results-rq1}) & \\ 

\quad{Tools Mentioned}             
    & 410 (45.96\%)  \\
\quad{Use Cases Mentioned}         
    & 325 (36.44\%) \\ 

\textbf{Perceptions of LLM Tools} (\S~\ref{sec:results-rq2}) & \\

\quad{Opinions Shared}    
    & 406 (45.52\%) \\
\quad{LLM Factors} 
    & 459 (51.46\%) \\ 
    
\textbf{Implications of LLM Adoption} (\S~\ref{sec:results-rq3}) &  \\

\quad{LLM Adoption}                 
    & 373 (41.82\%)  \\
\quad{Vision for the Future}       
    & 276 (30.94\%)  \\ 

\bottomrule
\end{tabular}
}

\label{tab:post-distribution}
\end{table}
\begin{table}[thbp]
\centering
\caption{\textbf{Comprehensive List of Reported LLM Tools} -- 
\textmd{\small grouped by general-purpose and security-specific platforms, along with their observed frequencies.}}
\footnotesize
\renewcommand{\arraystretch}{1.13}
\setlength{\tabcolsep}{4pt}

\resizebox{0.5\textwidth}{!}{
\begin{tabular}{p{1.9cm} >{\centering\arraybackslash}p{1.0cm} p{7.25cm}}
\toprule
\textbf{Tool Name} & \textbf{Freq.} & \textbf{Description} \\
\midrule

\multicolumn{3}{@{}l}{\textbf{General-Purpose LLM Tools}} \\

ChatGPT
& 90
& A general-purpose conversational LLM developed by OpenAI \\

MS Copilot
& 30
& A generative AI assistant integrated across Microsoft products \\

Claude
& 9
& A conversational LLM developed by Anthropic \\

Gemini
& 6
& A multimodal LLM developed by Google \\

Llama
& 5
& A family of open-source large language models released by Meta \\

Perplexity
& 3
& A generative AI-powered web search assistant \\

NotebookLM
& 3
& A research and note-taking online tool powered by Google Gemini \\

Grok
& 2
& A conversational LLM developed by xAI \\

Amazon Q
& 1
& Amazon’s enterprise generative AI assistant for AWS customers \\

\midrule

\multicolumn{3}{@{}l}{\textbf{Security-Specific LLM Tools}} \\

Security Copilot
& 40
& Microsoft’s LLM-powered agentic security automation platform \\

Dropzone AI
& 10
& Autonomous LLM-powered agentic ``AI SOC Analyst'' platform \\

Intezer
& 8
& Autonomous LLM-powered agentic ``AI SOC Analyst'' platform \\

Cortex XSIAM
& 6
& Extended security intelligence automation management platform \\

Prophet Security
& 4
& Autonomous LLM-powered agentic ``AI SOC Analyst'' platform \\

Purple AI
& 4
& Autonomous LLM-powered agentic ``AI SOC Analyst'' platform \\

CMD Zero
& 3
& Autonomous \& AI-assisted cyber investigation platform \\

Abnormal
& 3
& AI-native platform for human behavior security \\

Google SecOps
& 3
& Google’s intelligence-driven security operations platform \\

Darktrace
& 3
& AI-powered proactive platform for enterprise security \\

Torq Socrates
& 2
& Autonomous LLM-powered agentic ``AI SOC Analyst'' platform \\

Qevlar AI
& 2
& Autonomous LLM-powered agentic ``AI SOC Analyst'' platform \\

Arcanna AI
& 2
& Trustworthy agentic ``AI SOC Analyst'' platform \\

Vectra AI
& 2
& AI-powered platform for network, identity, and cloud security \\

WhiterabbitNeo
& 2
& Cybersecurity model built for offensive reasoning \\

D3 Morpheus
& 1
& Autonomous LLM-powered agentic ``AI SOC Analyst'' platform \\

TandemTrace
& 1
& Autonomous LLM-powered agentic ``AI SOC Analyst'' platform \\

Radiant Security
& 1
& Autonomous LLM-powered agentic ``AI SOC Analyst'' platform \\

Charlotte AI
& 1
& Autonomous LLM-powered agentic ``AI SOC Analyst'' platform \\

Exaforce
& 1
& Autonomous LLM-powered agentic ``AI SOC Analyst'' platform \\

7ai
& 1
& Autonomous LLM-powered agentic ``AI SOC Analyst'' platform \\

Rapid7
& 1
& AI-powered MDR platform for business resilience \\

Whistic
& 1
& AI-first platform for comprehensive third-party risk management \\

Splunk ES
& 1
& AI-powered threat detection, investigation, and response platform \\

SIRP
& 1
& Autonomous LLM-powered agentic ``AI SOC Analyst'' platform \\

HackerAI 
& 1
& LLM-powered penetration testing platform \\

XBOW
& 1
& LLM-powered penetration testing platform \\

Nebula AI
& 1
& LLM-powered penetration testing platform \\

Gradient Cyber
& 1
& AI-assisted MXDR designed for mid-market organizations \\

ReliaQuest
& 1
& Autonomous LLM-powered agentic ``AI SOC Analyst'' platform \\

\bottomrule
\end{tabular}%
}

\label{tab:tools-list}
\end{table}

\begin{table}[thbp]
\centering
\caption{\textbf{Practitioner Sentiment Across LLM Tools.}}
\footnotesize
\renewcommand{\arraystretch}{1.05}
\setlength{\tabcolsep}{4pt}

\begin{tabularx}{\columnwidth}{@{}X c c@{}}
\toprule
\textbf{LLM Tools} & \textbf{\# Positive} & \textbf{\# Negative} \\
\midrule

\textbf{General-Purpose Tools}
& \textbf{88 (50.57\%)}
& \textbf{86 (49.43\%)} \\

\quad Unnamed (LLMs, GenAI, etc.)
& 36 (42.35\%)
& 49 (57.65\%) \\

\quad ChatGPT and OpenAI models
& 38 (58.46\%)
& 27 (41.54\%) \\

\quad Microsoft Copilot
& 10 (50.00\%)
& 10 (50.00\%) \\

\quad Claude
& 4 (66.67\%)
& 2 (33.33\%) \\

\quad Gemini
& 0 (0.00\%)
& 5 (100.00\%) \\

\midrule

\textbf{Security-Specific Tools}
& \textbf{56 (52.83\%)}
& \textbf{50 (47.17\%)} \\

\quad Unnamed (AI SOC Analyst, Agentic SOC)
& 37 (59.68\%)
& 25 (40.32\%) \\

\quad Microsoft Security Copilot
& 8 (24.24\%)
& 25 (75.76\%) \\

\quad Intezer
& 4 (100.00\%)
& 0 (0.00\%) \\

\quad Dropzone
& 3 (100.00\%)
& 0 (0.00\%) \\

\quad Cortex XSIAM
& 3 (100.00\%)
& 0 (0.00\%) \\

\bottomrule
\end{tabularx}

\vspace{2pt}
\raggedright
\footnotesize{Counts and percentages are reported row-wise and represent the relative proportion of opinions within each LLM tool category. 
Only tools with three or more total opinion mentions are included.}

\label{tab:tool-opinions}
\end{table}
\header{Opinions on LLM Tools}
To further examine whether practitioner sentiment differed across categories of LLM tools, we analyzed positive and negative opinions associated with general-purpose and security-specific tools (Table~\ref{tab:tool-opinions}).
Overall, discussions surrounding the broader categories exhibited relatively balanced sentiment distributions, with no statistically significant differences observed at the category level ($\chi^2(1)=0.06$, $p=0.81$). 
However, tool-specific patterns reveal more nuanced perceptions. 
Among general-purpose systems, generic references to LLMs and GenAI systems were more commonly associated with negative opinions, whereas discussions of ChatGPT were comparatively more positive. 
Notably, while overall sentiment toward security-specific platforms remained slightly more positive, discussions surrounding Microsoft Security Copilot were substantially more negative than positive, suggesting a disconnect between broader excitement around AI SOC systems and practitioner experiences with specific enterprise deployments. 
Together, these findings suggest that practitioner sentiment toward LLM tools is shaped less by broad tool categories and more by hands-on experiences with specific platforms and their perceived operational value within SOC workflows.

\header{Relative Prevalence of Reported Use Cases}
As briefly discussed in Section~\ref{subsec:tasks}, to understand the relative prevalence across the six types of use cases, we conducted pairwise two-sample tests of proportions with a Bonferroni correction (Table~\ref{tab:usecase}).
We observed that Triage and IR were significantly more prevalent than all other reported uses ($\chi^2(1) \geq 16.92$, all $p < 0.001$), followed by writing-oriented tasks, including scripting, query generation, and reporting.
Statistically, scripting ($n${=}$88$) and reporting ($n${=}$84$) did not differ significantly from one another ($\chi^2(1) = 0.07, p = 0.79$), but were both significantly higher than threat analysis and knowledge support ($\chi^2(1) \geq 7.74, p < 0.05$).
Lastly, the discussions around threat analysis and knowledge support did not differ significantly ($\chi^2(1) = 0.00, p = 1.00$), but both were greater than misc.

\header{Correlating Tools with Use Cases}
To examine whether the distribution of discussed use cases differed between general-purpose and security-specific LLM tools, we conducted a chi-square test of independence followed by Bonferroni-corrected post-hoc analysis of adjusted residuals~(Table~\ref{tab:usecase}).
Due to its small sample size, the ``Training, Compliance, Others'' category was excluded from this analysis. 
The chi-square test revealed a significant association between use case category and tool category ($\chi^2(4)=58.289$, $p<0.01$).
Post-hoc analysis showed that triage and IR discussions were significantly overrepresented among security-specific tools ($z$=$6.83$, $p<0.001$). 
In contrast, scripting and query support ($z$=$3.57$, $p<0.01$) and reporting and documentation ($z$=$3.22$, $p<0.05$) discussions were significantly overrepresented among general-purpose tools.
No statistically significant differences were observed for knowledge support or threat analysis.

\header{Statistical Analysis of Factors and Perceptions}
To understand whether sentiment toward LLM tools differed across the discussed factors, we constructed a contingency table capturing positive versus negative mentions for each factor~(Table~\ref{tab:factor-def}), and conducted a chi-square test of independence.
The test revealed a significant association between factor type and sentiment ($\chi^2(5) = 172.21, p < 0.001$), indicating that practitioners’ perceptions differed significantly across factors. 
Post-hoc tests with adjusted residuals find that comments around LLMs' capabilities
$(z = 8.11, p < .001)$, and efficiency $(z = 6.38, p < .001)$ were significantly more likely to be positive than negative, while
reliability $(z = -6.77, p < .001)$, 
security and privacy $(z =$ \text{--}5.41, $p < .001)$, 
level of independent autonomy $(z = -4.94, p < .001)$, 
and cost $(z = -4.02, p < .001)$ 
were significantly more likely to be negative. 

\header{Temporal Analysis of LLM Adoption Stages}
To examine whether practitioner discussions surrounding LLM adoption changed over time, we conducted chi-square tests of independence between adjacent temporal intervals: p1 vs p2 and p2 vs p3 (Table~\ref{tab:adoption}), and measured delta values to represent proportional changes in the relative prevalence of each adoption stage between adjacent periods.
No statistically significant difference was observed between the first two phases, p1 and p2 ($\chi^2(2)=0.44$, $p=0.802$). 
Relative proportions suggested that discussions during the earlier periods remained similarly centered around evaluating and exploring LLM tools. 
In contrast, a significant shift was observed between p2 and p3 ($\chi^2(2)=13.20$, $p=0.001$). 
Post-hoc analysis using Bonferroni-corrected adjusted residuals revealed a transition in practitioner discourse toward active operational use: exploratory discussions became significantly underrepresented in p3 ($z=-3.58$, $p=0.002$), whereas discussions describing active use became significantly overrepresented ($z=3.07$, $p=0.013$). 
No statistically significant temporal differences were observed for discussions expressing non-adoption.

\subsection{Impact of LLMs on Security Workforce}
\label{subsec:future-jobs}

Building on observed adoption patterns and barriers (\S~\ref{subsec:adoption}, \S~\ref{subsec:future-caution}), practitioners’ discussions also reflected on the implications of LLMs for the SOC workforce. Despite vendor claims of autonomy and replacement, practitioner discussions revealed a more nuanced view of how LLM tools reshape roles, responsibilities, and skill demands within SOCs.

\header{Where LLMs Can Meaningfully Augment Humans} 
Across practitioner discussions ($n${=}$28$), a widely shared belief was that if LLM tools were to replace any SOC role, {L1 responsibilities are the most vulnerable.}
Practitioners argued that L1 positions were already being reduced prior to the LLM hype and that the widespread adoption of LLMs is likely to accelerate this trend.
For example, as one practitioner, P250, explained how their company \pquote{let go of all eight L1 SOC members, because the SOAR playbooks handled almost everything, and phishing, the only task manually reviewed, was later handed off to an AI tool.}
Moreover, several \posts frame L1 workflows as \pquote{button-clicking} (P916), \pquote{brain-dead work} (P244), \pquote{barely a security role} (P237), or \pquote{simply processing routine tasks shown by the SIEM} (P304), 
justifying that LLM tools need not human-level reasoning to affect L1 staffing.
Building on this reasoning, while some practitioners predicted substantial reductions in entry-level positions, \pquote{AI could impact headcount by 15–20\%} (P227).

\header{Where Human Expertise Remains Critical}
Despite concerns about the future of L1 roles, practitioners overwhelmingly rejected the idea that SOCs are close to becoming fully autonomous.
Practitioner discussions consistently emphasized that even as LLM tools become deeply embedded in workflows, {human oversight remains indispensable}.

\textit{High-skill SOC Responsibilities:}
Across 20 posts, practitioners emphasized that many high-skill SOC responsibilities, including digital forensics and IR, threat hunting, and penetration testing (P711, P367, P539, P927), typically handled by L2–L3 teams, inherently require humans.
As P279 emphasized, \pquote{Any task that requires complex reasoning, logical synthesis and judgment, I see people having a strong presence in handling.} 
This perspective was further reinforced by stressing the criticality of human expertise: \pquote{Even the most advanced AI tools are ineffective without a skilled security team to implement them or without strong executive backing from a CISO (or equivalent)} (P631).

\textit{LLM Supervision and Governance:}
Another set of posts ($n${=}$18$) highlighted that {with LLMs in the picture, humans are needed more than ever}.
Practitioners consistently pointed out that organizations will still need humans to \pquote{supervise and verify LLM outputs} (P005, P243, P401, P390).
As P243 summarized, \pquote{Even with LLMs, there will still be a need for certain levels of verification, which in itself could be an L1 responsibility.}
Practitioners further stressed that \pquote{certain responsibilities, such as governance and compliance, cannot be delegated to AI, as doing so risks allowing the system to effectively oversee itself.} (P241).
They also argued that within SOC, AI cannot operate without human-provided context.
As P354 explained, even someone skilled at prompting requires substantial prior experience to guide the LLM: 
\newpage
\begin{quote}
    \textit{If a company hires someone who can generate solutions through effective prompting, that alone does not make them a replacement for skilled analysts. Meaningful use of LLMs still requires domain knowledge and expertise of an experienced analyst.}
\end{quote}

\textit{Accountability:}
Lastly, across a small set of posts ($n${=}$6$), practitioners highlighted {humans will still be needed for accountability}, arguing that organizations cannot solely rely on LLM tools for decisions that may have legal, regulatory, or financial consequences (P048, P237, P285).
This was especially evident for incident response workflows. 
As P347 noted, \pquote{Unsure what a breach response would look like if the `AI employee' overlooked anything and it caused harm to people... some `human' will ultimately need to be held responsible.}

\header{How Analysts Must Adapt}
Beyond the debate between replacement and augmentation, a few \posts ($n${=}$11$) highlighted that early adopters of LLM tools are already integrating them into daily workflows, and therefore avoiding LLMs or dismissing their relevance should be an untenable stance: \pquote{We would be naïve to overlook the scale and speed of change underway, no one can predict with certainty how the field will evolve} (P275).
Consequently, practitioners offered explicit advice to peers on how to navigate this shift.

\textit{Developing AI Literacy:}
A commonly repeated suggestion was {developing AI literacy}, followed by the sentiment that \pquote{AI itself is not the reason for job-threat; rather, it is the peers who learn to use it effectively} (P248). 
Several, including P246, P818, and P843, described \pquote{AI as a bell that cannot be unrung,} (P246) stressing that analysts who fail to build fluency will be the first to fall behind as organizations increasingly seek people who can work confidently with AI-enabled tooling. 
P292 contextualized this urgency by pointing out the unprecedented pace of LLM adoption, noting that, \pquote{only a few technologies in recent history have achieved such rapid global familiarity and enterprise adoption.}

\textit{More Depth in Security Reasoning and Response:}
Another category of advice emphasized the importance of {strengthening fundamentals} and {continuous upskilling}.
As one practitioner, P304, advised, \pquote{To anyone aspiring, make upskilling a part of your DNA. Go beyond simply responding to alerts to actually understand why detections fire and how systems operate.}
Others also underscored the ongoing importance of \pquote{hands-on experience with traditional SOC tools such as SIEM and EDR} (P034), as well as familiarity with emerging areas such as AI security threats, e.g., ISO 27090~(P271).

\end{appendices}

\end{document}